\documentclass[manuscript]{aastex}
\usepackage{amsmath}
\usepackage{index}
\usepackage{color}
\usepackage{aas_macros}
\font\manual=manfnt at 7pt \def\dbend{\hbox{\raise0.9ex\hbox{\manual\char127\hspace{0.6em}}}}

\newcommand\Ion[2]{\ensuremath{\mathrm{#1\,\scriptstyle #2}}}

\newcounter{INTERNALionstage}
\providecommand{\ion}[2]{% replace the aastex version
  \setcounter{INTERNALionstage}{#2}%
  \Ion{#1}{\Roman{INTERNALionstage}}}

\def\gtsim{\mathrel{\hbox{\rlap{\hbox{\lower4pt\hbox{$\sim$}}}\hbox{$>$}}}}
\def\lesssim{\mathrel{\hbox{\rlap{\hbox{\lower4pt\hbox{$\sim$}}}\hbox{$<$}}}}

\def\cm{{\rm\thinspace cm}}

\def\pcc{{\rm\thinspace cm^{-3}}}

\def\pscm{\mbox{$\cm^{-2}\,$}}

\def\pscm{\mbox{$\cm^{-2}\,$}}

\DeclareMathAlphabet{\vib}{OML}{cmm}{m}{it}
\begin{document}

\title{Atomic data for \ion{S}{2} - Toward Better Diagnostics of Chemical 
Evolution in High-redshift Galaxies}
\author{
Romas Kisielius\altaffilmark{1}, 
Varsha P. Kulkarni\altaffilmark{2}, 
Gary J. Ferland \altaffilmark{3},
Pavel Bogdanovich\altaffilmark{1},
Matt L. Lykins\altaffilmark{3}
}

\altaffiltext{1}{Institute of Theoretical Physics and Astronomy, Vilnius 
University, A. Go{\v s}tauto 12, LT-01108, Lithuania}
\altaffiltext{2}{Department of Physics and Astronomy, University of South 
Carolina, Columbia, SC 29208, USA}
\altaffiltext{3}{Department of Physics and Astronomy, University of Kentucky, 
Lexington, KY 40506, USA}

\begin{abstract}
Absorption-line spectroscopy is a powerful tool used to estimate element 
abundances in the nearby as well as distant universe. The accuracy of the 
abundances thus derived is, naturally, limited by the accuracy of the atomic 
data assumed for the spectral lines. We have recently started a project to 
perform the new extensive atomic data calculations used for 
optical/UV spectral lines in the plasma modeling code Cloudy using 
state-of-the-art quantal calculations. Here we demonstrate our approach by 
focussing on \ion{S}{2}, an ion used to estimate metallicities for Milky Way 
interstellar clouds as well as distant damped Lyman-alpha (DLA) and sub-DLA 
absorber galaxies detected in the spectra of quasars and gamma-ray 
bursts (GRBs). We report new extensive calculations of a large number of 
energy levels of \ion{S}{2}, and the line strengths of the resulting radiative 
transitions.
Our calculations are based on the configuration interaction approach within 
a numerical Hartree-Fock framework, and utilize both non-ralativistic and
quasirelativistic one-electron radial orbitals.  The results of these new 
atomic calculations are then incorporated into Cloudy and applied to a lab 
plasma, and a typical DLA, for illustrative purposes. The new results imply 
relatively  modest changes ($\approx 0.04$ dex) to the metallicities estimated 
from \ion{S}{2} in past studies. These results will be readily applicable to 
other studies of \ion{S}{2} in the Milky Way and other galaxies.\\

\end{abstract}


\section{Introduction}

A powerful tool for studying distant galaxies is provided by absorption lines in
the spectra of quasars superposed by foreground galaxies along the sightline, 
which are sampled simply by gas cross-section, independent of their brightness. 
Damped Lyman-alpha absorbers (DLAs; neutral hydrogen column densities 
$N_{\rm H I}  \ge 2 \times 10^{20}$ cm$^{-2}$) and sub-DLAs 
($10^{19} \le N_{\rm H I} < 2 \times 10^{20}$ cm$^{-2}$) are especially useful 
for this purpose. These   are the primary neutral gas reservoir for star 
formation at redshifts $0 <z < 5$ [e.g., \cite{Storrie00}; \cite{Peroux05}; 
\cite{Prochaska05}]. Over the past decade, DLAs toward gamma-ray burst (GRB) 
afterglows have also emerged as a powerful probe of distant galaxies 
[e.g., \cite{Savaglio03}; \cite{Chen05}; \cite{Prochaska07}; \cite{Fynbo09}].

The elemental compositions of DLAs/sub-DLAs offer highly sensitive tracers of 
the chemical evolution of galaxies  [e.g., \cite{Pettini97}; \cite{KF02}; 
\cite{Prochaska03}; \cite{Kulkarni07}; \cite{Prochaska07}; \cite{Peroux08};  
\cite{Meiring09}; \cite{Cooke11}; \cite{Rafelski12}; \cite{Som13}]. 
Element abundances in the absorbers are determined from optical/UV atomic 
resonance lines. Volatile elements such as N, O, P, S, Ar, Zn are not strongly 
condensed on interstellar dust grains; so their gas-phase abundances can give 
their total (gas + solid phase) abundances. In practice, for reasons such as 
wavelength coverage and availability of suitable lines, S and Zn 
have emerged as the most commonly used metallicity indicators used for DLAs. 
The common ionization stage of S seen in cool interstellar clouds and DLAs 
is \ion{S}{2}. \ion{S}{2} has several absorption lines that can be used to 
obtain reliable column densities despite their usual presence in the Lyman-alpha 
forest.

The accuracy of the element abundances and physical properties inferred from 
them depends crucially on the quality of the atomic data used.  By far, 
the commonly used atomic data reference for DLA spectral analysis is \cite{Morton03} 
 [see, e.g., \cite{Battisti12}; \cite{Rafelski12}; \cite{Kulkarni12}; 
\cite{Guimaraes12}; \cite{Jorgenson13}; \cite{Som13}]. For some important 
transitions and ions, the oscillator strengths listed in Morton (2003) have 
relatively large uncertainties [as listed on the NIST Atomic Spectra Database
\cite{asd13}], while for some transitions, oscillator strengths are not 
available at all. In some cases, even more recent values obtained since 
\cite{Morton03} have low accuracy grades listed in the NIST
 database. Limitations in atomic data can compromise our ability
to read the messages received from high-redshift galaxies. 
To produce new reliable atomic data for commonly used astrophysical ions, 
we have recently started a collaborative study that 
brings together atomic physics, plasma simulations, and observational 
spectroscopy. Our goals are to assess the quality of the existing atomic data, 
to improve accuracy of the data that were designated low accuracies, 
to incorporate them into our widely used plasma simulation code Cloudy, and 
apply it to existing/new observations of high-redshift galaxies such as 
DLAs/sub-DLAs. Here we provide an early illustration of our approach by 
focussing on \ion{S}{2}, an ion of great importance for DLA element 
abundance studies. 

Being a volatile element, S does not condense easily on interstellar dust 
grains. In the Milky Way, S shows a depletion of $< 0.1$ dex in cool as well as 
warm interstellar clouds [e.g., \cite{Savage96}; but see also \cite{Jenkins09} 
for the suggestion that the true depletion of S could be larger  in the presence
of ionization effects]. The relatively low depletion makes S ideal for 
estimating metallicity from gas-phase abundance measurements. Moreover, S is a 
fairly abundant element, so its  absorption lines are easily detectable (more 
easily detectable than the lines of Zn, another nearly undepleted element). 
Especially important among the S ions is \ion{S}{2}, which is the dominant ion 
in DLAs. \ion{S}{2} has a number of absorption lines, especially a triplet at 
$\lambda \, \lambda$ 1250.6, 1253.8, 1259.5 {\AA}, that are strong enough to be 
detectable. The weakest of these lines can be relatively unsaturated, giving 
reasonably accurate column densities (although the stronger lines can be 
saturated). Studies of DLAs, especially at $z \gtrsim 2$ often use these three 
\ion{S}{2} lines to derive [S/H], and adopt that as the gas-phase 
metallicity. Studies of the Milky Way interstellar gas have also often adopted 
S as a metallicity indicator. For all of these calculations, it is naturally 
important to use accurate atomic data for \ion{S}{2}. 

\cite{Morton03} lists the oscillator strengths for the above-mentioned three 
\ion{S}{2} lines to be 0.00543, 0.0109, and 0.0166, respectively, but the 
uncertainties in these values are classified as ``C" grade, i.e. at the level 
of about 25$\%$ as per the NIST database. Thus, the uncertainties in the 
metallicity introduced by the uncertainty in the oscillator strength 
($\sim 0.1$ dex) are far larger than those often quoted from the measurement 
uncertainties in high-resolution data (typically $\lesssim 0.05$ dex). 
\ion{S}{2} also has absorption lines at 906.9, 910.5, and 912.7 {\AA}, but they 
are too close to the Hydrogen Lyman edge and much stronger and hence likely 
saturated. Other \ion{S}{2} lines at 943.0, 947.0, 1021.3, and 1021.5 {\AA} are 
also listed in \cite{Morton03}, but without oscillator strength estimates.

With the desire to assess the atomic data accuracy, we undertook new calculations of the 
oscillator strengths for all \ion{S}{2} electric dipole, magnetic dipole, and 
electric quadrupole transitions. Section 2 describes these new calculations and 
compares them to previous estimates. Section 3 describes the incorporation of 
these calculations into Cloudy, and section 4 discusses the implications for 
DLA abundance studies. 

\section{Calculations of new atomic data}

A broad study of energy levels, oscillator strengths and transition 
probabilities for the levels of some low configurations of \ion{S}{2} was 
performed by \citet{cff05}. The authors used multiconfiguration 
Hartree-Fock method (MCHF) with relativistic effects included in the Breit-Pauli
approximation (BP) in their study. They have determined all possible allowed E1
and many forbidden (E2, M1) transitions for the states under consideration and 
determined level lifetimes and splittings. As a  first step in these calculations,
{\sl ab initio} wavefunctions were obtained, and in the next step, the diagonal 
energies of $LS$ blocks were adjusted in order to get better agreement of the 
energies of $LS$ terms with the observed values. We refer to the results of these 
calculations as MCHF$_{05}$.

A following study by \citet{cff06} considered energy levels, lifetimes and 
transition probabilities for several sequences, including the \ion{S}{2} ion as
a member of P-like sequence. These authors used several theoretical methods, 
such as non-orthogonal spline configuration interaction, multiconfiguration 
Hartree-Fock, and multiconfiguration Dirac-Hartree-Fock. Transitions between the 
computed levels were reported for allowed E1 and some forbidden (M1, M2, E2, E3)
transitions. The MCHF wavefunction expansion adopted in this work was very 
similar to that of \citet{cff05}, but there were no term-corrections included 
in the Hamiltonian matrix. In the comparisons given below, we will use the 
results of the \citet{cff06} calculations as consistent configuration 
interaction (CI) {\sl ab initio} calculation data and refer to them as MCHF$_{06}$.

Recently, \cite{Tayal10}  reported new calculations for transition 
probabilities and electron-impact collision strengths for the astrophysically 
important lines in \ion{S}{2}. The multiconfiguration Hartree-Fock method with 
term-dependent non-orthogonal orbitals was employed for accurate representation 
of the target wavefunctions. Relativistic corrections were included in the 
Breit-Pauli approximation. Their close-coupling expansion included 70 bound 
levels of \ion{S}{2} covering all possible terms of the ground $3s^23p^3$ 
configuration and singly excited $3s3p^4$, $3s^23p^23d$, $3s^23p^24s$, and 
$3s^23p^24p$ configurations. This approach made it possible to achieve a more 
accurate description of both energy levels and oscillator strengths with a 
relatively small CI expansion compared to that of more traditional methods with 
an orthogonal set of one-electron orbitals where large CI expansions are 
necessary. According to \cite{Tayal10}, the accuracy of their calculations is 
comparable with the accuracy of the Breit-Pauli MCHF calculations by \cite{cff05}
discussed earlier. In further references, we refer to the calculations of 
\cite{Tayal10} as MCHF$_{\mathrm{TD}}$.

A systematic study of forbidden M1 and E2 transitions for \ion{P}{1}, 
\ion{S}{2}, \ion{Cl}{3}, and \ion{Ar}{4} was reported by \cite{mcdf99}. They
applied multiconfiguration Dirac-Fock (MCDF) wavefunctions of different 
sizes to check for the convergence of results. These authors concluded that
the convergence and reasonable agreement of their calculations with previously
determined results could be achieved only after a large number of valence- and 
core-excited configurations were included in their multiconfiguration 
wavefunction expansion. 

The above-mentioned works are the most systematic and complex theoretical studies
of allowed and forbidden transitions in \ion{S}{2} so far. They provide more
reliable line data compared to the earlier relativistic results of \cite{mend82}
and \cite{keenan93}. A recent compilation by \cite{Podobedova09} has tabulated
more than 6000 allowed and forbidden lines of \ion{S}{1} to \ion{S}{15}.
This study provides a critical evaluation of recent theoretical values for
transition rates, and also includes energy level values that are primarily 
experimental, taken from the NIST compilation by \cite{asd13}. 
Specifically for \ion{S}{2}, \cite{Podobedova09} list transition probabilities 
for allowed E1 transitions and forbidden M1 and E2 transitions.  In this 
compilation, the estimated uncertainties of theoretical values in many E1 
transitions exceed 25\% while M1 and E2 transitions are given better accuracy.

In the present work, we employ two different approximations for the calculation of 
\ion{S}{2} transition rates. In the first one, a multiconfiguration Hartree-Fock
method is adopted where relativistic corrections are included in the Breit-Pauli 
approximation. It resembles the method used in \cite{cff06}, but there are several 
significant differences. We adopt transformed radial orbitals (TRO) 
\citep{tro04, tro05} in order to efficiently include electron correlation 
corrections  caused by excited configurations with higher principal quantum 
numbers $n > 4$. In the current work, the transformed one-electron radial orbitals
$P_{\mathrm{TRO}}(nl|r)$ have two variational parameters, an integer and even 
$k$ and a positive $B$:  
\begin{align}
\label{tro}
P_{\mathrm{TRO}}(nl|r) = &
N ( r^{l-l_0+k} \exp(-Br)P(n_0l_0|r) \nonumber \\ 
- & \sum_{n^{\prime} < n} P(n^{\prime}l|r) \int_{0}^{\infty} 
P(n^{\prime}l|r^{\prime})r^{\prime\;(l-l_0+k)} \exp(-Br^{\prime})
P(n_0l_0|r^{\prime})dr^{\prime} ) .
\end{align}
Here the factor $N$ ensures the normalization of the determined TROs, the first
term in the parenthesis performs the transformation of RO based on the one-electron
radial orbital $P(n_0l_0|r)$ from the set of investigated configurations, and the 
second term ensures their orthogonality. The parameters $k$ and $B$ are chosen to gain
the maximum of the energy correlation correction. In the current calculation, 
TROs were determined for the configurations with the outer electron having the
principal quantum number $5 \leq n \leq 7$ and all allowed values of the orbital 
quantum number $l$.

The second rather significant difference is in the selection of the 
configurations included in the CI wave function expansion. Instead of simply 
including higher-$n$ excited configurations, we follow the procedure described 
by \cite{pbrk01} and remove those configurations (admixed configurations) within 
the CI wavefunction expansion of the investigated configuration (adjusted 
configuration) which have the mean weight 
${\bar W}_{\mathrm{PT}} < 1 \times 10^{-8}$. 

The parameter ${\bar W}_{\mathrm{PT}}$ is determined in the second order of perturbation 
theory (PT):
\begin{equation}
\label{eq-w}
{\bar W}_{\mathrm{PT}}(K_0,K^{\prime}) = 
\frac{\sum_{TLST^{\prime}}(2L+1)(2S+1) 
\langle K_0TLS \Vert H \Vert K^{\prime}T^{\prime}LS \rangle ^2}
{g(K_0) \left( {\bar E}(K^{\prime}) - {\bar E}(K_0) \right)^2},
\end{equation}
where $\langle K_0TLS \Vert H \Vert K^{\prime}T^{\prime}LS \rangle$ is a 
Hamiltonian matrix element  for the interaction between the adjusted $K_0$ and 
the admixed $K^{\prime}$ configuration $LS$-terms, $g$ is the statistical weight
of the configuration $K_0$, and ${\bar E}$ are the averaged energies of the 
configurations.
This method, paired with the methods from \cite{pbrkam02}, makes it possible to  
significantly reduce the size of the Hamiltonian matrices.

In the CI approximation, our methods allow for the use of two kinds of radial 
orbitals describing the electrons of adjusted configurations which are applied
for the transformations (\ref{tro}). In the case of \ion{S}{2}, we have electrons 
with the principal quantum number $n \leq 4$. Traditionally, the solutions
of the standard Hartree-Fock equations are utilized for this purpose, see, e.g.
\citet{pbrkim03, vjfe20, vjfe22, rkpb09}. We use the notation 
CI$_{\mathrm{HF+TRO}}$ to denote the results obtained for this approximation. 
Relativistic corrections are included in the Breit-Pauli approximation, as in 
the MCHF calculations.

In order to partially account for relativistic corrections at the stage when 
the one-electron radial orbitals are determined, we developed a new method which 
solves the quasirelativistic (QR) Hartree-Fock equations. The quasirelativistic 
radial orbitals, obtained after solving the QR equations, are applied to 
determine the one-electron wavefunctions of admixed configurations 
and further to calculate the transformed radial orbitals given by Eq.~\ref{tro}. 
A consequent inclusion of correlation effects is achieved by the same method as 
in the case of the afore-mentioned non-relativistic Hartree-Fock radial orbitals. 
To determine the energy levels, the Breit-Pauli approximation is applied as in 
the CI$_{\mathrm{HF+TRO}}$ calculations. We use the CI$_{\mathrm{QR+TRO}}$ 
notation for this method. Furthermore, we must mention that our QR method 
differs significantly from the more traditional quasirelativistic method of
\cite{cowan}. A more detailed description of the applied QR approximation can be 
found elsewhere [see, e.g., \cite{pbor06, pbor07, pbor08, pbrk12, pbrk13}].

\section{Accuracy of atomic data sets}

One the main tasks of the present paper is to assess the accuracy of the
spectroscopic data used in modeling the \ion{S}{2} emission or absorption
spectra. In order to do that, we compare our results, both in
the CI$_{\mathrm{HF+TRO}}$ and CI$_{\mathrm{QR+TRO}}$ approximations, with
those from \cite{cff06} and  \cite{Tayal10} calculations.

\subsection{Energy levels and wavelengths}

Table~\ref{tab_lev} presents the energy levels of \ion{S}{2}. The results of our 
calculations obtained with the two methods are compared with the experimental 
data and with the values calculated by \cite{cff06} and \cite{Tayal10}. One can 
see from the mean-square deviations ($MSD$) provided at the bottom of 
Table~\ref{tab_lev} that the data of \cite{Tayal10} agree with the experimental 
values better than with both sets of our data or those of \cite{cff06}, although 
the differences among calculated values are rather small. The better accuracy of 
the \cite{Tayal10} data can be explained by the fact that those authors used the 
term-dependent non-orthogonal radial orbitals to calculate energy levels and 
radiative transition parameters. Such an approximation determines the 
eigenvalues for each $LS$-term separately instead of optimizing a complete set 
of terms.

%%%%%%%%%%%%%%%%%%%%%%%%%%%%%%%%%%%%%%%%%%%%%%%%
\renewcommand{\baselinestretch}{0.75}
\begin{deluxetable}{rlrrrrrr}
%\tabletypesize{\scriptsize}
\tabletypesize{\footnotesize}
\tablecolumns{8} 
\tablewidth{0pt}
\tablecaption{
\label{tab_lev}
Comparison of calculated \ion{S}{2} energy levels with experimental data. 
} 
\tablehead{ 
\colhead{$N$} &
\colhead{State} & 
\colhead{$J$} &
\colhead{CI$_{\mathrm{HF+TRO}}$} & 
\colhead{CI$_{\mathrm{QR+TRO}}$} & 
\colhead{Exp} &
\colhead{MCHF$_{06}$} &
\colhead{MCHF$_{\mathrm{TD}}$} 
}
\startdata 
  1& $3p^3  $ \; $^4S $&   1.5&      0&      0&      0&      0&      0\\    
  2& $3p^3  $ \; $^2D $&   1.5&  15292&  15320&  14853&  15282&  14880\\    
  3& $3p^3  $ \; $^2D $&   2.5&  15307&  15335&  14885&  15311&  14905\\    
  4& $3p^3  $ \; $^2P $&   0.5&  25407&  25549&  24525&  24817&  24632\\    
  5& $3p^3  $ \; $^2P $&   1.5&  25431&  25574&  24572&  24848&  24656\\    
  6& $3s3p^4$ \; $^4P $&   2.5&  77787&  78926&  79395&  78468&  79405\\    
  7& $3s3p^4$ \; $^4P $&   1.5&  78106&  79249&  79757&  78763&  79704\\    
  8& $3s3p^4$ \; $^4P $&   0.5&  78290&  79434&  79963&  78929&  79873\\    
  9& $3s3p^4$ \; $^2D $&   2.5&  97281&  98322&  97919&  97500&  97875\\    
 10& $3s3p^4$ \; $^2D $&   1.5&  97284&  98324&  97891&  97481&  97899\\    
 11& $3p^23d$ \; $^2P $&   1.5& 105241& 105702& 105599& 105444& 105530\\    
 12& $3p^23d$ \; $^2P $&   0.5& 105599& 106062& 106044& 105846& 105933\\    
 13& $3p^24s$ \; $^4P $&   0.5& 109399& 109344& 109561& 108939& 109635\\    
 14& $3p^24s$ \; $^4P $&   1.5& 109600& 109548& 109832& 109182& 109877\\    
 15& $3p^24s$ \; $^4P $&   2.5& 109927& 109880& 110269& 109570& 110264\\    
 16& $3p^23d$ \; $^4F $&   1.5& 110068& 110275& 110177& 110174& 110216\\    
 17& $3p^23d$ \; $^4F $&   2.5& 110179& 110387& 110313& 110297& 110337\\   
 18& $3p^23d$ \; $^4F $&   3.5& 110336& 110547& 110509& 110473& 110514\\    
 19& $3p^23d$ \; $^4F $&   4.5& 110543& 110757& 110767& 110705& 110748\\    
 20& $3p^24s$ \; $^2P $&   0.5& 113060& 113029& 112938& 112487& 113055\\    
 21& $3p^24s$ \; $^2P $&   1.5& 113449& 113422& 113462& 112952& 113514\\    
 22& $3p^23d$ \; $^4D $&   0.5& 113864& 114082& 114162& 113986& 114144\\    
 23& $3p^23d$ \; $^4D $&   1.5& 113896& 114115& 114201& 114020& 114176\\    
 24& $3p^23d$ \; $^4D $&   2.5& 113940& 114160& 114231& 114064& 114216\\    
 25& $3p^23d$ \; $^4D $&   3.5& 113996& 114216& 114279& 114120& 114265\\    
 26& $3p^23d$ \; $^2F $&   2.5& 114942& 115190& 114804& 115018& 114853\\    
 27& $3p^23d$ \; $^2F $&   3.5& 115310& 115563& 115286& 115437& 115281\\    
 28& $3s3p^4$ \; $^2S $&   0.5& 119175& 119904& 119784& 119887& 119862\\    
 29& $3p^24s$ \; $^2D $&   1.5& 121656& 121612& 121529& 121230& 121499\\    
 30& $3p^24s$ \; $^2D $&   2.5& 121658& 121614& 121530& 121230& 121499\\    
 31& $3p^23d$ \; $^2G $&   3.5& 127320& 127489& 127127& 127771& 127161\\    
 32& $3p^23d$ \; $^2G $&   4.5& 127332& 127501& 127128& 127773& 127161\\    
 33& $3p^23d$ \; $^4P $&   2.5& 131737& 132161& 130602& 131100& 130775\\    
 34& $3p^23d$ \; $^4P $&   1.5& 131913& 132343& 130819& 131292& 130960\\    
 35& $3p^23d$ \; $^4P $&   0.5& 132019& 132452& 130949& 131406& 131073\\    
 36& $3p^23d$ \; $^2D $&   1.5& 134648& 135023& 133361& 133915& 133469\\    
 37& $3p^23d$ \; $^2D $&   2.5& 135065& 135454& 133815& 134346& 133864\\    
 38& $3p^24s$ \; $^2S $&   0.5& 136559& 136753& 136329& 136026& 136315\\    
 39& $3p^23d$ \; $^2P $&   0.5& 140742& 140939& 139845& 140485& 139881\\    
 40& $3p^23d$ \; $^2P $&   1.5& 140826& 141021& 140017& 140667& 140034\\    
 41& $3p^23d$ \; $^2F $&   3.5& 141879& 142499& 138527& 139950& 138639\\    
 42& $3p^23d$ \; $^2F $&   2.5& 141896& 142516& 138509& 139956& 138614\\    
 43& $3p^23d$ \; $^2D $&   2.5& 145843& 146418& 144009& 144982& 144308\\    
 44& $3p^23d$ \; $^2D $&   1.5& 146011& 146599& 144142& 145138& 144422\\    
 45& $3s3p^4$ \; $^2P $&   1.5& 152020& 153750& 145506& 145933& 145631\\    
 46& $3s3p^4$ \; $^2P $&   0.5& 152326& 153955& 145878& 146418& 145946\\    
 47& $3p^23d$ \; $^2D $&   2.5& 152340& 153207& 148887& 150450& 149067\\    
 48& $3p^23d$ \; $^2D $&   1.5& 152491& 153257& 148901& 150614& 149075\\    
 49& $3p^23d$ \; $^2S $&   0.5& 154834& 155791& 151652& 153365& 151745\\
\tableline
$MSD$&                 &      & 1710  &   2071&    -- &   698 &    94 \\  
\enddata 

\tablecomments {
CI$_{\mathrm{HF+TRO}}$ - our HF data;
CI$_{\mathrm{QR+TRO}}$ - our quasirelativistic data; 
Exp - experimental values;
MCHF$_{06}$ - data from \cite{cff06};
MCHF$_{\mathrm{TD}}$ - data from \cite{Tayal10}.
}
\end{deluxetable}

Nevertheless, some deviations in energy level values (consequently, in 
transition wavelength values) are not significant, and they do not exeed $3\%$. 
Moreover, these discrepancies can be overcome by using the experimental 
wavelengths, which are well-known for the most important \ion{S}{2} lines, 
for the line identification or by using the experimental energy differences to 
determine the ``corrected'' oscillator strengths or radiative transition 
probabilities [see \cite{Verner96}].

\subsection{E1 lines}

We have determined radiative transition parameters for the lines arising from 
the transitions among the levels of $3s^23p^3$, $3s3p^4$, $3p^23d$, and $3p^24s$
configurations of \ion{S}{2}. The electric dipole, electric qudrupole and 
magnetic dipole transitions were considered. The main target of the present work
is not only to determine high-accuracy radiative transition data but also to
evaluate the accuracy of the calculations and their suitability for use in the 
plasma modeling code Cloudy.

\begin{figure}[!th]
\includegraphics[scale=0.6]{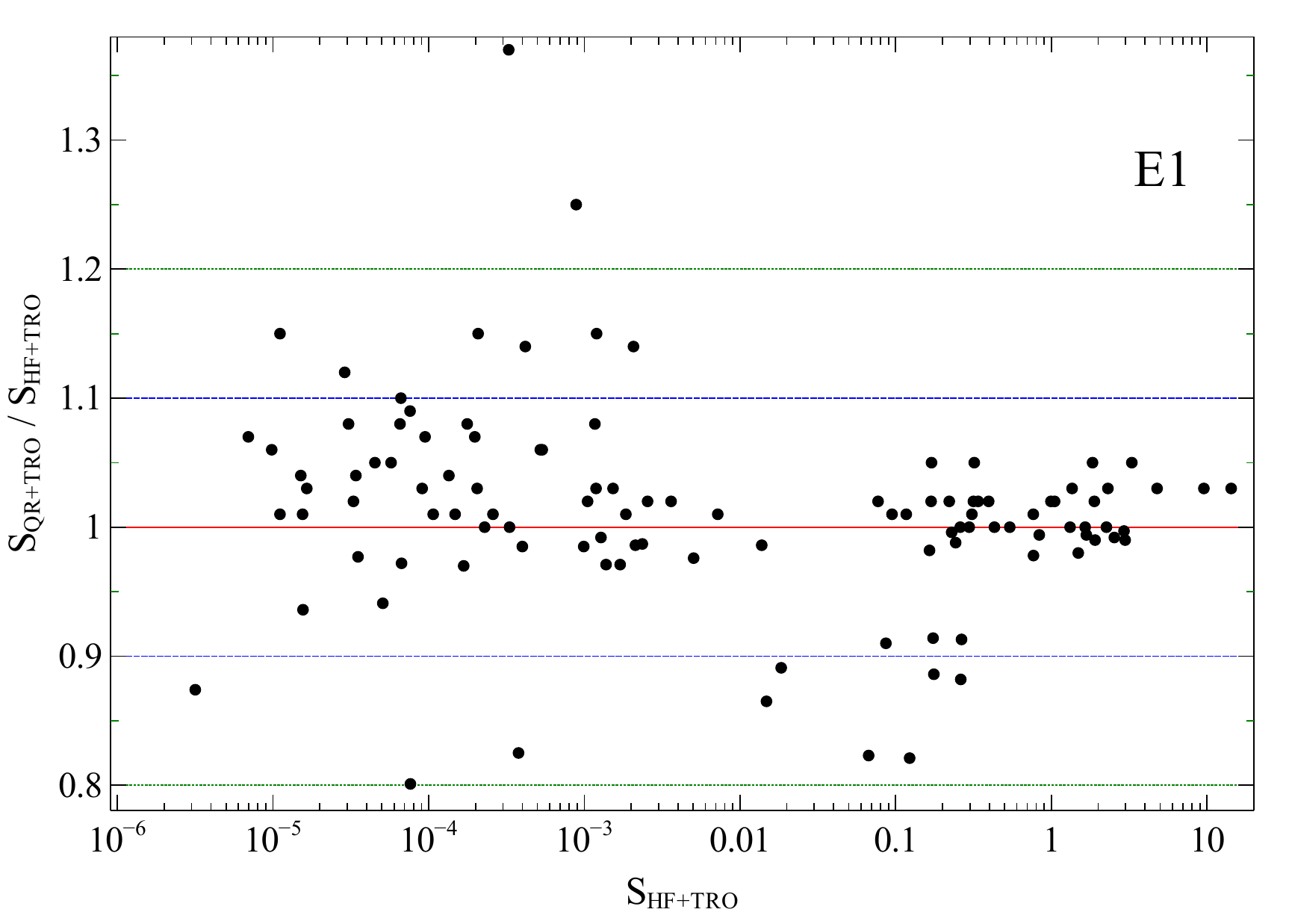}
\caption{ 
Comparison of the E1 transition line strengths $S$ determined in the 
CI$_{\mathrm{HF+TRO}}$ and CI$_{\mathrm{QR+TRO}}$ approaches for \ion{S}{2}.
\label{fig_e1qr}
}
\end{figure}

\begin{figure}[!th]
\includegraphics[scale=0.6]{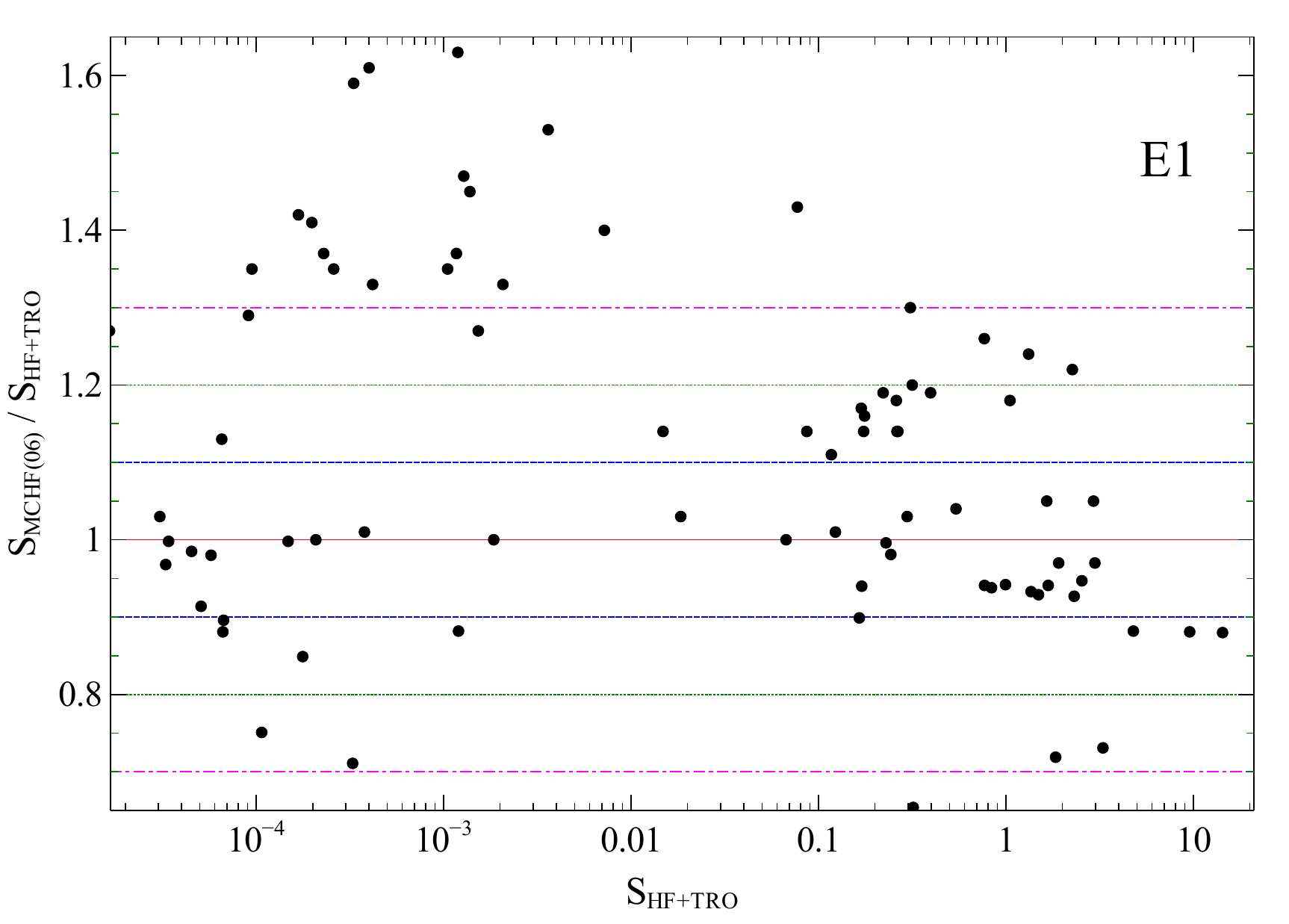}
\caption{ 
Comparison of the E1 transition line strengths $S$ determined in the 
CI$_{\mathrm{HF+TRO}}$ and MCHF$_{06}$ \citep{cff06} approaches for \ion{S}{2}.
\label{fig_e1cff}
}
\end{figure}

\clearpage
\begin{figure}[!th]
\includegraphics[scale=0.6]{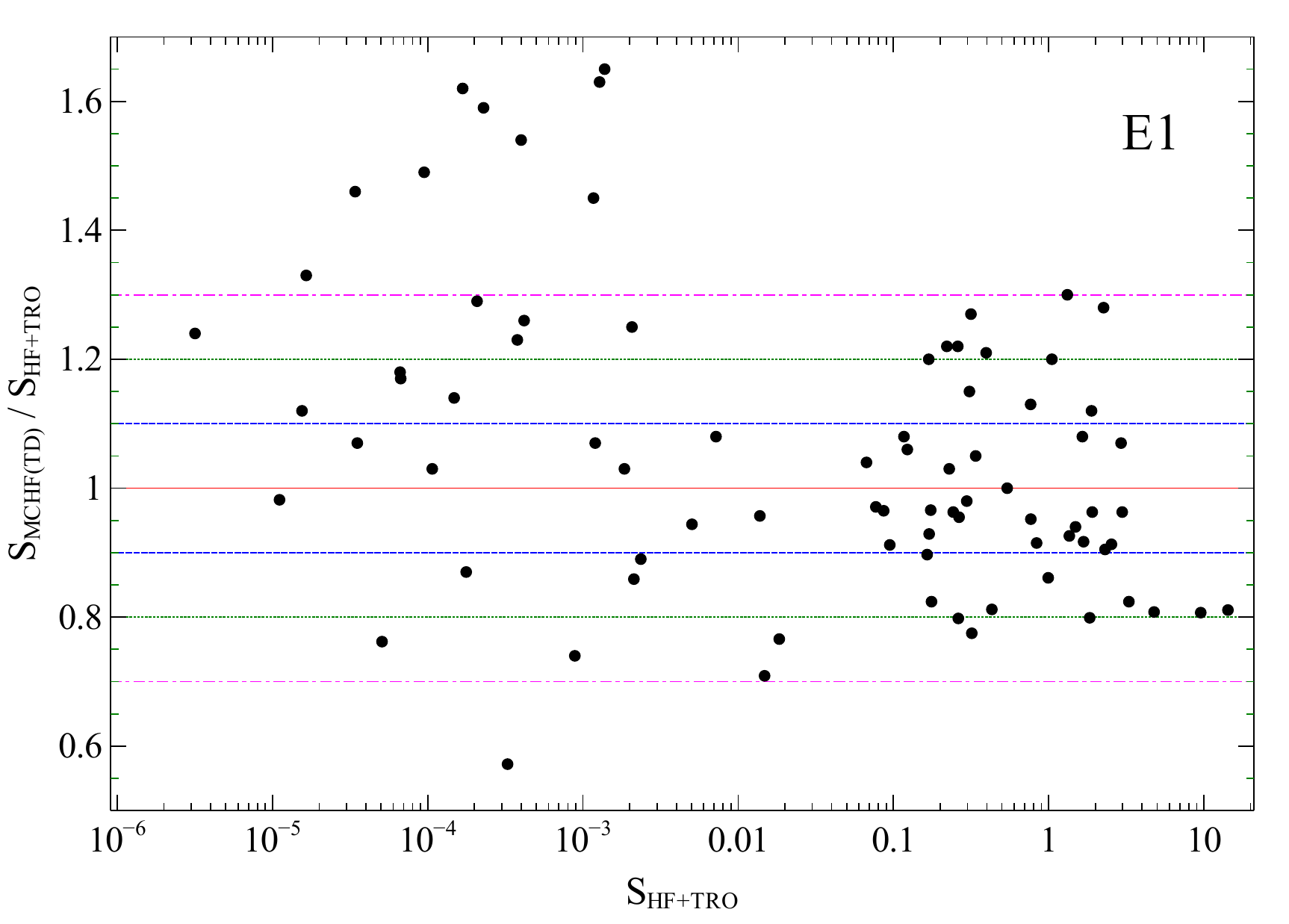}
\caption{ 
Comparison of the E1 transition line strengths $S$ determined in the 
CI$_{\mathrm{HF+TRO}}$ and MCHF$_{\mathrm{TD}}$ \citep{Tayal10} approaches 
for \ion{S}{2}.
\label{fig_e1bs}
}
\end{figure}

One of the ways to assess the accuracy of our results is to compare 
them with other available data. For this purpose we make a comparison of the
E1 transition line strengths $S$. Transition line strengths $S$ are chosen, 
because they do not depend directly on the transition energy difference, unlike
the oscillator strengths $f$ or transition probabilities $A$. As it is mentioned
before, the accuracy of $f$-values and $A$-values can be increased by using 
theoretically calculated $S$-values and the experimental transition energy or
the wavelength $\lambda$.

Figure \ref{fig_e1qr} compares our calculated results determined 
using two different approximations. The CI$_{\mathrm{HF+TRO}}$ approach utilizes 
non-relativistic radial orbitals while the CI$_{\mathrm{QR+TRO}}$ approach 
employs quasirelativistic radial orbitals. These two approaches adopt the same 
configuration interaction method, involving the transformed radial orbitals, 
to deal with the correlation effects. 
As one can see, agreement of the results is rather nice, especially for the
strongest lines. Within 2 orders of magnitude, the discrepancies do not exceed
$10\%$, except for 3 lines. Within 4 orders of magnitude, the most discrepancies
are within  $10\%$, but there are 5 more lines with the discrepancies ranging
from  $10\%$ to $20\%$. Extending comparison to 6 orders of magnitude, situation 
does not change substantially, and the most lines agree within $10\%$. 
Nevertheless, there are some lines where discrepancies exceed $20\%$ or even 
$30\%$.

Comparison of our CI$_{\mathrm{HF+TRO}}$ calculation results with the data from 
\cite{cff06} is given in Fig.~\ref{fig_e1cff}. As one can see, the agreement of 
these data is worse in Fig.~\ref{fig_e1cff}, as compared to Fig.~\ref{fig_e1qr}. 
This is caused by the use of different CI expansion bases in our and \cite{cff06} 
calculations. In general, the discrepancies for most of the strong lines are 
within $20\%$ except for the few lines with the discrepancies larger than 
that, when transition line strengths within 3 orders of magnitude are considered. 
One can see some larger deviations, exceeding $30\%$ in this comparison but the 
number of such lines is small. For weaker lines, there is a large number of 
lines with discrepancies exceeding $30\%$.

In Fig. \ref{fig_e1bs}, a similar comparison of our CI$_{\mathrm{HF+TRO}}$ results 
with the data from \cite{Tayal10} is presented. Here again, the discrepancies do 
not exceed $30\%$ and in most cases are below $20\%$ for the line strength range
of 4 orders of magnitude. There are just 5 lines with discrepancies exceeding 
$25\%$. When weaker lines are considered, agreement of data resulting from 
different approximations is worse, and in some cases, the discrepancies can 
reach $50\%$ or even more.

We conclude that our two sets of calculations for the radiative transition 
data agree very well between themselves for the strongest transitions, within 6 
orders of magnitude of the largest line strengths. This proves that relativistic 
corrections are included adequately in our CI$_{\mathrm{HF+TRO}}$ approximation. 
For the weakest lines, where transitions purely depend on configuration interaction 
effects, this close agreement breaks down even if the CI basis remains the same. 
Comparison with the data from \cite{cff06} and \cite{Tayal10} demonstrates, 
that, for the strongest lines, deviations are within $20\%$ to $30\%$  for the 
line strengths varying by 4 orders of magnitude from the strongest. Agreement 
for many weaker lines is worse and exceeds $30\%$, as the above-mentioned 
studies apply different CI bases compared to our calculations.

\renewcommand{\baselinestretch}{1.1}
\begin{deluxetable}{lrlrrrrrr}
\tabletypesize{\footnotesize}
\tablecolumns{9} 
\tablewidth{0pt}
\tablecaption{
\label{tab_acc}
Comparison of calculated transition rates $A$ (s$^{-1}$) for some E1 
transitions to the ground state  $3p^3\;^4S_{3/2}$. 
}
\tablehead{ 
\colhead{Excited state} &
\colhead{$\lambda_{\mathrm{Exp}}$(\AA)} &
\colhead{$ag$} &
\colhead{CI$_{\mathrm{HF+TRO}}$} & 
\colhead{CI$_{\mathrm{QR+TRO}}$} & 
\colhead{MCHF$_{\mathrm{TD}}$} &
\colhead{MCHF$_{06}$}&
\colhead{MCHF$_{05}$}&
\colhead{$gf^{corr}$}
}
\startdata 
$3s3p^4\,\, ^4P_{5/2}     $ &1259.518& {\it D+}& 4.48E+7& 4.02E+7& 4.27E+7& 4.92E+7& 5.10E+7& 6.27E-2\\ 
$3s3p^4\,\, ^4P_{3/2}     $ &1253.805& {\it D+}& 4.48E+7& 4.00E+7& 4.30E+7& 4.93E+7& 5.10E+7& 4.14E-2\\ 
$3s3p^4\,\, ^4P_{1/2}     $ &1250.578& {\it D} & 4.48E+7& 4.00E+7& 4.32E+7& 4.94E+7& 5.11E+7& 2.06E-2\\ 
$3s3p^4\,\, ^2D_{5/2}     $ &1021.254& {\it ng}& 3.41E+4& 3.47E+4& 3.50E+4& 2.51E+4& 2.61E+4& 3.17E-5\\ 
$3s3p^4\,\, ^2D_{3/2}     $ &1021.539& {\it ng}& 5.28E+3& 5.40E+3& 5.17E+3& 3.54E+3& 3.67E+3& 3.28E-6\\ 
$3p^2(^3P)3d\,\, ^2P_{3/2}$ & 946.978& {\it ul}& 2.42E+2& 1.49E+3& 4.24E+3& 2.68E+3& 2.10E+3& 8.02E-7\\ 
$3p^2(^3P)3d\,\, ^2P_{1/2}$ & 943.003& {\it ul}& 9.87E+1& 1.08E+3& 9.45E+2& 1.20E+3& 7.65E+2& 2.88E-7\\ 
$3p^2(^3P)4s\,\, ^4P_{1/2}$ & 912.735& {\it D+}& 1.12E+9& 1.10E+9& 1.02E+9& 1.03E+9& 1.05E+9& 2.78E-1\\ 
$3p^2(^3P)4s\,\, ^4P_{3/2}$ & 910.484& {\it C} & 1.13E+9& 1.11E+9& 1.03E+9& 1.04E+9& 1.06E+9& 5.60E-1\\ 
$3p^2(^3P)4s\,\, ^4P_{5/2}$ & 906.885& {\it C} & 1.15E+9& 1.13E+9& 1.05E+9& 1.07E+9& 1.08E+9& 8.49E-1\\ 
\enddata

\tablecomments {
Transition wavelengths $\lambda_{\mathrm{exp}}$ are experimental values. 
CI$_{\mathrm{HF+TRO}}$ - our MCHF data adjusted for the experimental transition 
energies;
CI$_{\mathrm{QR+TRO}}$ - our quasirelativistic Hartree-Fock data; 
MCHF$_{\mathrm{TD}}$ - data from \cite{Tayal10};
MCHF$_{06}$ - data from \cite{cff06};
MCHF$_{05}$ - data from \cite{cff05}.
$gf^{corr}$ are ``corrected'' $gf$ values based on our calculations adjusted for
the experimental transition energies.
Accuracy grades $ag$ are taken from \cite{asd13}, 
{\it C} means assumed $25\%$ accuracy,
{\it D} means assumed $40\%$ accuracy,
{\it D+} means assumed $50\%$ accuracy,
{\it ng} means no accuracy grade is given, 
{\it ul} means lines are not listed in \cite{asd13}.
}
\end{deluxetable}

The particular \ion{S}{2} lines listed in Table 2 were chosen because they are 
all of the lines from \ion{S}{2}  listed by \cite{Morton03}, which is the 
commonly used reference in observational spectroscopy of quasar absorption 
systems and Galactic interstellar medium. The most commonly used lines are the 
triplet at 1250.578, 1253.805, 1259.518 {\AA}, as they have determinations of 
oscillator strengths and are relatively easy to observe being longward of the 
\ion{H}{1} Lyman-$\alpha$ transition, and often outside the red wing of the 
Lyman-$\alpha $ transition. The doublets at or below 912 {\AA} are difficult to 
observe due to their proximity to the \ion{H}{1} Lyman limit. In the case of 
DLA/sub-DLAs (and in general the Lyman-limit systems), absorption near the 
\ion{H}{1} Lyman limit renders the quasar flux to be nearly zero in this 
wavelength region.  Nevertheless, we list them in Table~\ref{tab_acc} for the 
sake of completeness. The doublets at 1021.254, 1021.539 {\AA} and at 943.003, 
946.978 {\AA} can in principle be observed with more ease, but have no 
measurements of oscillator strengths listed in \cite{Morton03}. From our 
calculations for these lines as well as those from other works presented in 
Table~\ref{tab_acc}, it is clear that observing these lines will be far more 
challenging. But the strongest of these lines (at 1021.539 {\AA}) may be 
possible to detect in the strongest absorbers. For example, in a 
solar-metallicity absorber with log $N_{\rm H I} = 22.0$, the S II 
$\lambda$ 1021.254 line would be expected to have a rest-frame equivalent width 
of $\approx 12$ m{\AA}.

In the last column of Table~\ref{tab_acc} we present the $gf$-values for these
lines. These values are derived from the line strengths $S$ calculated in the 
CI$_{\mathrm{HF+TRO}}$ approximation with the experimentally adjusted transition
energies. 
We give oscillator strengths as the product $gf$ due to its symmetry.  
We note that the emission $f_{u,l}$ and absorption $f_{l,u}$ values are related 
by  $g_u f_{u,l} = -g_l f_{l,u}$.
The lines originating from the excited $3p^2(^3P)3d\,\, ^2P_{3/2}$ and
$3p^2(^3P)3d\,\, ^2P_{1/2}$ levels (at 943.003, 946.978 {\AA}) are very weak. 
For them we find the largest discrepancies when different data sets are compared.
Unfortunately, our CI$_{\mathrm{HF+TRO}}$ results do not agree very well with 
other data, therefore $gf^{corr}$ values for these two lines were derived from
our CI$_{\mathrm{QR+TRO}}$ calculations which are in much better agreement with
the data from \cite{Tayal10} and \cite{cff06}.

\subsection{E2 and M1 lines}

\begin{figure}[!th]
\includegraphics[scale=0.6]{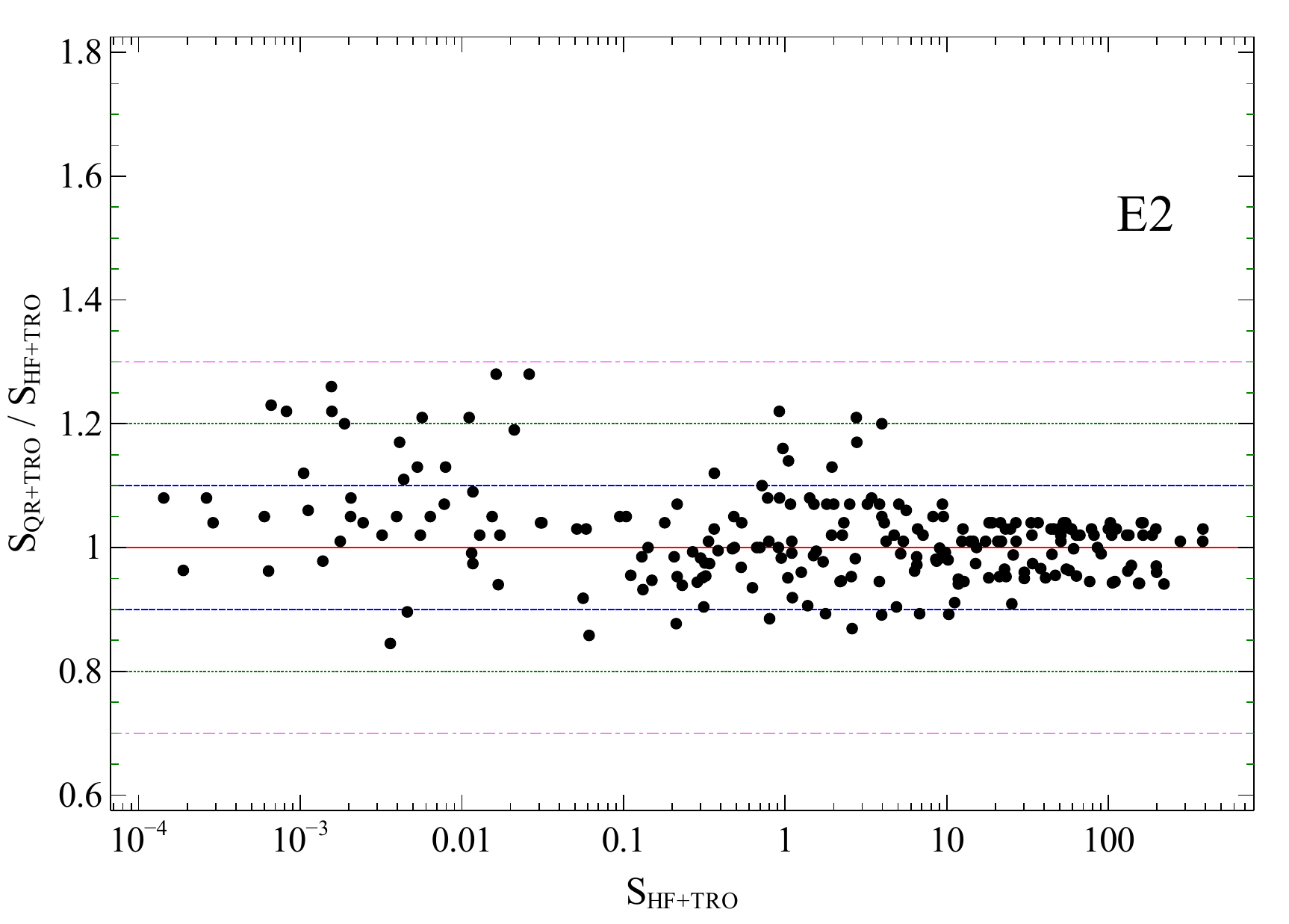}
\caption{ 
Comparison of the E2 transition line strengths $S$ determined in the 
CI$_{\mathrm{HF+TRO}}$ and CI$_{\mathrm{QR+TRO}}$ 
approaches for \ion{S}{2}.
\label{fig_e2qr}
}
\end{figure}

The line strengths $S$ of E2 transitions from our two sets of calculations are
compared in Fig.~\ref{fig_e2qr}. As one can see, the deviations do not exceed 
$20\%$, except for a few weaker lines. But even for these lines, the 
deviations are smaller than $30\%$. Moreover, the deviations for most of the 
lines are $< 10\%$. So here again, our two calculation methods, 
CI$_{\mathrm{HF+TRO}}$ and CI$_{\mathrm{QR+TRO}}$, produce very similar results.

\begin{figure}[!th]
\includegraphics[scale=0.6]{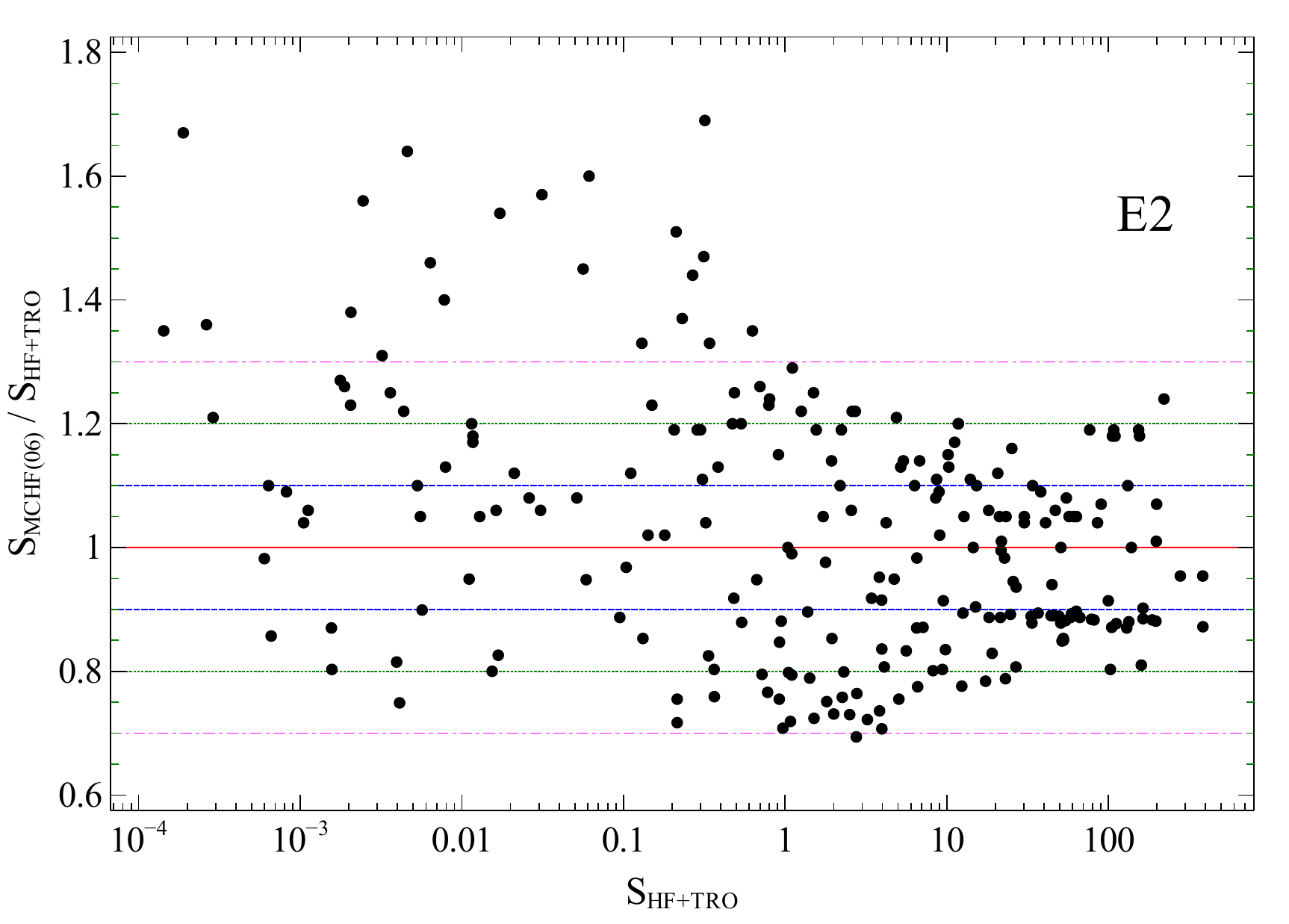}
\caption{ 
Comparison of the E2 transition line strengths $S$ determined in the
CI$_{\mathrm{HF+TRO}}$ and MCHF$_{06}$ \citep{cff06} approaches for \ion{S}{2}.
\label{fig_e2cff}
}
\end{figure}

A comparison of our CI$_{\mathrm{HF+TRO}}$ results with the data from \cite{cff06} 
is shown in Fig.~\ref{fig_e2cff}. It is evident that the scatter in the ratio
$S_{\mathrm{MCHF_{06}}}/S_{\mathrm{CI_{HF+TRO}}}$ is significantly larger 
compared to the scatter in the $S_{\mathrm{CI_{QR+TRO}}}/S_{\mathrm{CI_{HF+TRO}}}$ ratio. 
Nevertheless, the  lines within 2 orders of magnitude of the strongest have 
deviations smaller than $20\%$, while most of the remaining weaker lines have 
deviations smaller than $30\%$. We note that there are some rather strong lines 
($0.1 \leq S \leq 1$), which have deviations $> 30\%$ or even $> 40\%$. But the 
total number of lines having deviations larger than $30\%$ is around 20.

\begin{figure}[!th]
\includegraphics[scale=0.6]{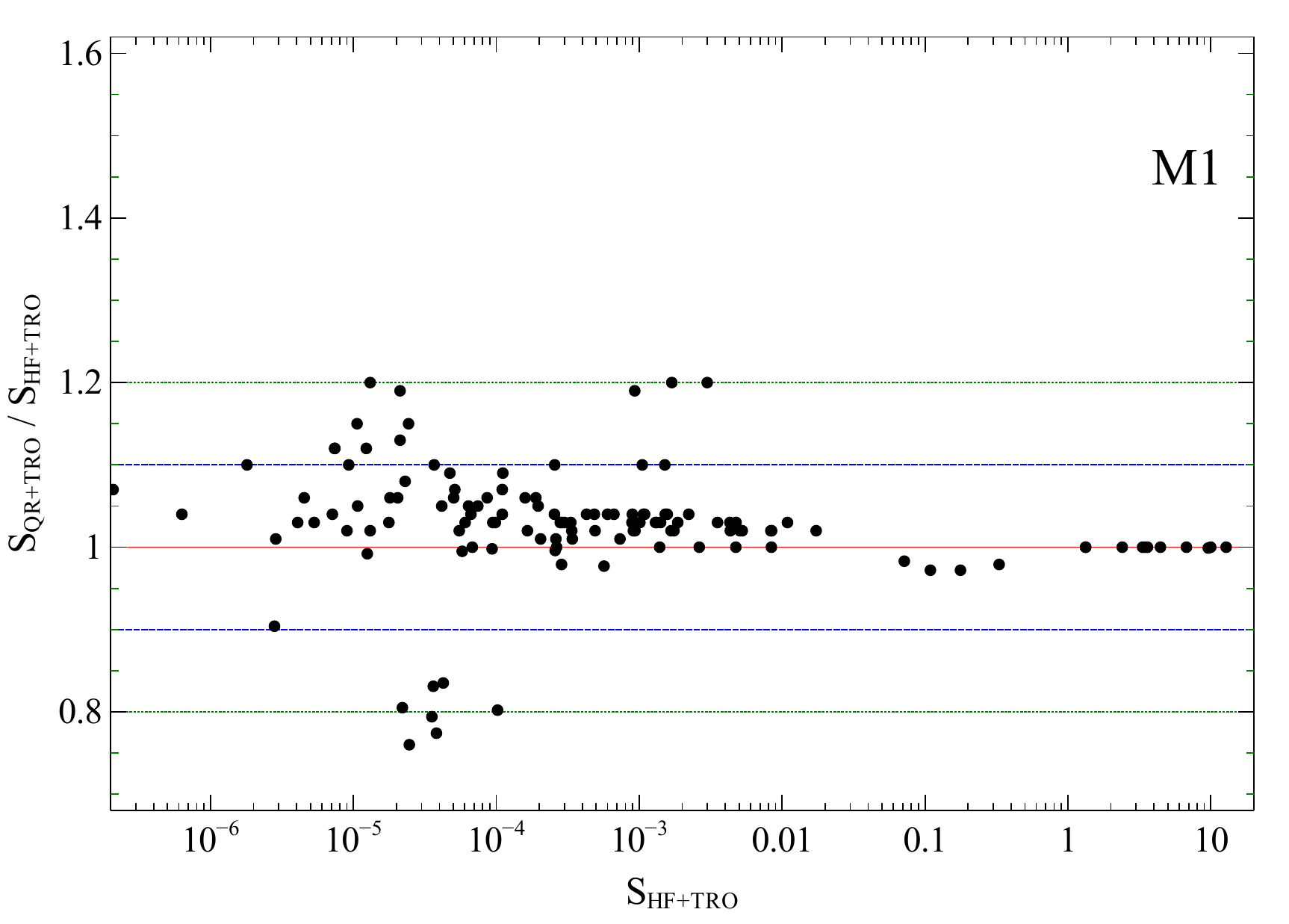}
\caption{ 
Comparison of the M1 transition line strengths $S$ determined in the
CI$_{\mathrm{HF+TRO}}$ and MCHF$_{06}$ \citep{cff06} approaches for \ion{S}{2}.
\label{fig_m1qr}
}
\end{figure}

\begin{figure}[!th]
\includegraphics[scale=0.6]{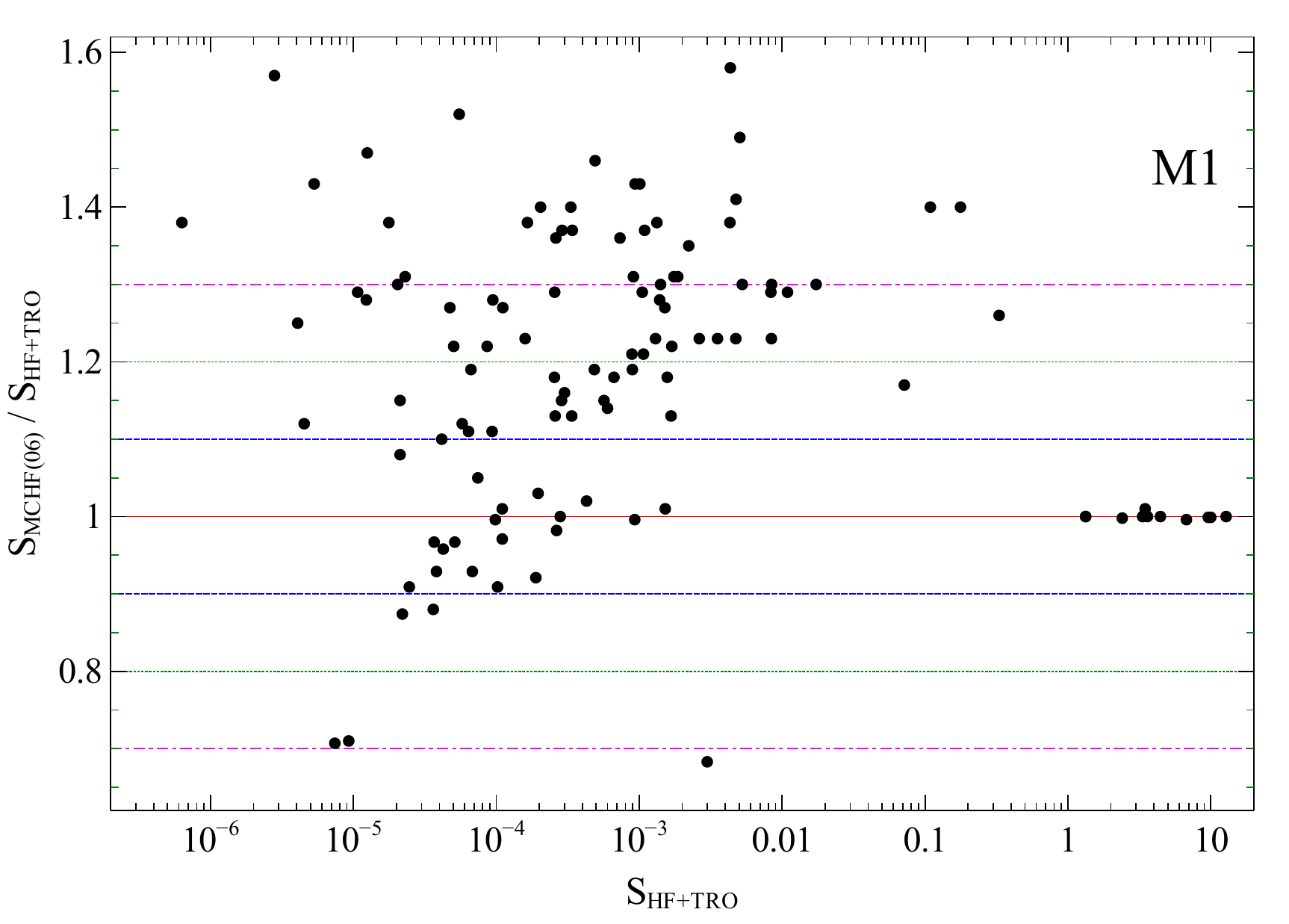}
\caption{ 
Comparison of the M1 transition line strengths $S$ determined in 
CI$_{\mathrm{HF+TRO}}$ and MCHF$_{06}$ \citep{cff06} approaches for \ion{S}{2}.
\cite{cff06} approaches for \ion{S}{2}.
\label{fig_m1cff}
}
\end{figure}

Figure 6 compares our calculated line
strengths for the M1 transitions. As one can see, the agreement is very good. 
For almost all the lines, the two values are within $10\%$ of one another, and only 3 rather weak lines
have deviations worse than $20\%$. A similar comparison with the data from 
\cite{cff06} is given in Fig.~\ref{fig_m1cff}. Unfortunately, in this case, 
the deviations are much larger, $> 20\%$ or 
even $30\%$ for a large fraction of the lines. Since the radiative M1 transition operator does not depend on the 
variable $r$, the main cause for the large deviations is that different CI expansion bases
are used in our calculation and in the calculation by \cite{cff06}.

\section{Cloudy calculations}

We have converted the new data to the {\sc stout} 
format, the data base introduced
in version 13 of Cloudy, the spectral synthesis code last described by
\citet{FerlandCloudy13}.
As described by \citet{Lykins13a, Lykins13b} and \citet{FerlandCloudy13}, 
Cloudy obtains much of its atomic and molecular
data from external files, making it far easier to update and modify the data.
For S II we combine our new calculations of the transition rates with NIST energy levels
and collision strengths given by \citet{Tayal10}.
We save line strengths $S$ rather than $A_{ul}$ or $gf$ because we use
experimental energies -- $S$, unlike the other two, does not depend directly on transition energy.

Table \ref{tab_tran} lists a few of the calculated transition line strengths $S$;
the majority are available in the on-line version. 
The level indices are from Table 1, and the experimental level energies given there can
be used to derive line wavelengths.
These data were
produced in the CI$_{\mathrm{HF+TRO}}$ approximation. Based on the listed data,
one can easily transform transition line strengths $S$ into the $gf$ values or 
the transition rates $A$ by using available experimental or calculated transition
energies. Further details are given in \citet{Lykins13b}.

%%%%%%%%%%%%%%%%%%%%%%%%%%%%%%%%%%%%%%%%%%%%%%%%
\renewcommand{\baselinestretch}{0.9}
\begin{deluxetable}{llrrl}
\tabletypesize{\footnotesize}
\tablecolumns{5} 
\tablewidth{0pt}
\tablecaption{
\label{tab_tran}
Transition line strengths $S$ (in a.u.) for \ion{S}{2} determined in the 
CI$_{\mathrm{HF+TRO}}$ appproximation. 
} 
\tablehead{ 
\colhead{Data} &
\colhead{Type} & 
\colhead{$N_l$} &
\colhead{$N_u$} &
\colhead{$S$}
}
\startdata 
S& E2&  1&   2& 5.54E$-$03\\
S& M1&  1&   2& 1.77E$-$05\\
S& E2&  1&   3& 1.29E$-$02\\
S& M1&  1&   3& 6.30E$-$07\\
S& E2&  1&   4& 2.25E$-$06\\
S& M1&  1&   4& 3.37E$-$04\\
S& E2&  1&   5& 3.00E$-$10\\
S& M1&  1&   5& 1.67E$-$03\\
S& E1&  1&   6& 2.65E$-$01\\
S& E1&  1&   7& 1.74E$-$01\\
S& M1&  2&   3& 2.40E$+$00\\
S& E2&  2&   4& 2.18E$+$01\\
S& M1&  2&   4& 2.63E$-$03\\
S& E2&  2&   5& 2.17E$+$01\\
S& M1&  2&   5& 8.41E$-$03\\
\enddata 

\tablecomments {
The first column describes the transition data type (S stands for 
line strengths $S$, A -- for transition rates $A$). 
The second columns describes line type, 
$N_l$ is for the lower level index, 
$N_u$ denotes the upper level index.}
\tablecomments {
(This table is available in its entirety in a machine-readable form in the online
journal. A portion is shown here for guidance regarding its form and content.)
}
\end{deluxetable}

The following sections show representative sulphur spectra and discuss
an application to DLAs.

\subsection{Pure-S$^+$ emission spectra}

\citet{Lykins13a} describe our calculation of gas in collisional ionization equilibrium.
We show two representative spectra, of absorption and emission spectra, in this section.

\begin{figure}
\includegraphics[scale=1.05]{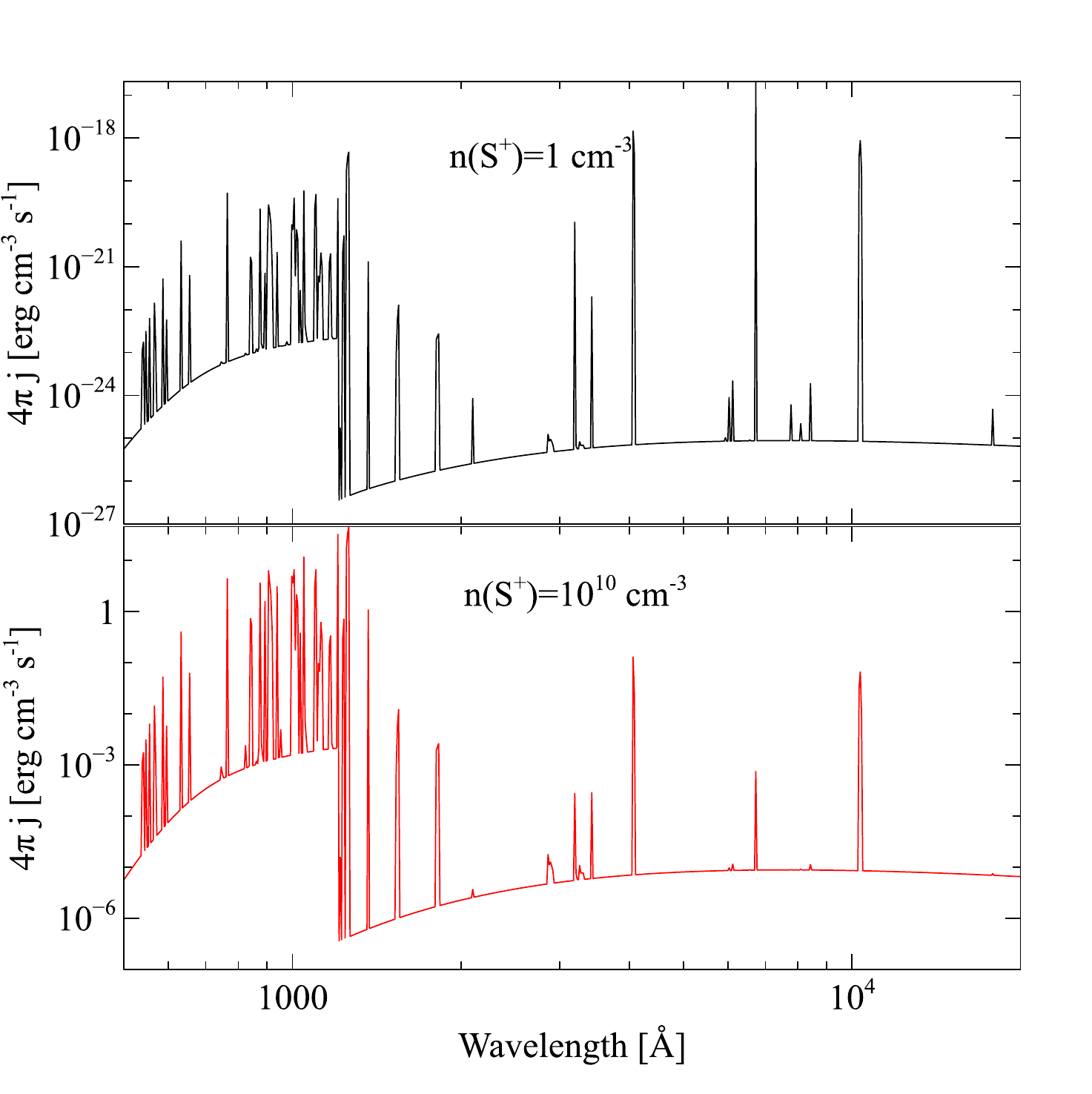}
\caption{
The \ion{S}{2} emission spectra of a unit cell of pure-S$^{+}$ gas
in coronal equilibrium at a temperature of $2\times 10^4$~K.
Two S$^{+}$ densities, $1 \pcc$ and $10^{10} \pcc$ are shown.
The low density produces a spectrum that peaks in the optical/NIR while
the high density case emits mainly in the FUV.
\label{fig_emi}
}
\end{figure}

Two emission spectra of a pure-S gas in coronal equilibrium at $T = 2 \times10^4$ K, 
the temperature where the fraction of S$^+$ peaks, is shown in Figure \ref{fig_emi}.  
Both simulations have a unit volume ($1 \pcc$) of gas but have different densities,  
$1 \pcc$ and $10^{10} \pcc$. 
The lower density is in the low-density limit and the spectrum would be characteristic
of any gas with density $n_e \lesssim 10^3 \pcc$.
As expected, the denser gas is $\sim 10^{20}$ times more emissive.
The higher density is characteristic of quasar emission-line regions \citep{AGN3}.
The prominent feature at 1197\AA\ is the S$^+ \rightarrow \mathrm{S}^0$  
radiative recombination continuum.
Many hundreds of S II lines are present, providing valuable diagnostics of the gas conditions.

The strongest lines in the lower density, upper panel of Figure \ref{fig_emi}, are 
the optical forbidden [\ion{S}{2}] $\lambda \lambda 6730.82, 6716.44 $ doublet, a density indicator in nebulae
\citep{AGN3}.
The next strongest lines are 
the optical [\ion{S}{2}] $\lambda \lambda 4068.60 , 4076.35$ doublet,
which, combined with the previous pair, are a temperature indicator.
Multiplets in the NIR at
$\lambda \lambda 10\,320,
10\,336,
10\,287,
10\,370$, and in the FUV at
$\lambda \lambda 1259.52,
1253.81,
1250.58$,
are also strong.

The optical and NIR forbidden lines are collisionally suppressed 
in the denser gas shown in the lower panel.
The strongest lines in this case are in the FUV, at
$\lambda \lambda 1259.52,
1253.81,
1250.58$
multiplet followed by
$\lambda \lambda 1204.32, 
1204.27$.
These lines are allowed and are optically thick if the S$^+$ column density is large enough.

\subsection{Pure-S$^+$  absorption spectra}

\begin{figure}[!th]
\includegraphics[scale=1.05]{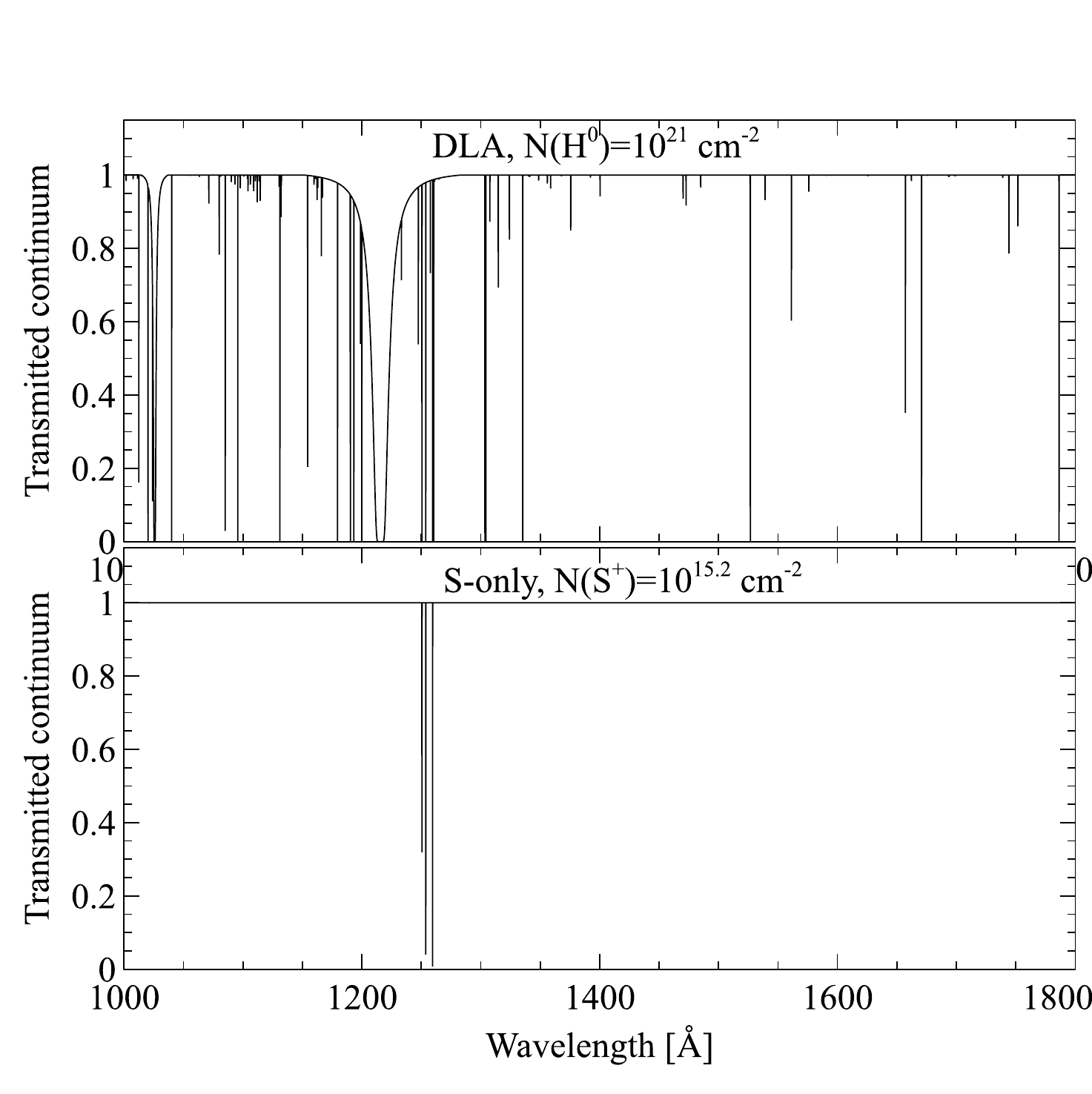}
\caption{ 
\ion{S}{2} absorption plot.  The upper panel shows a full DLA spectrum,
with absorption lines of various elements shown relative to a normalized continuum.
The strongest line is the damped Ly$\alpha$ line.
The lower panel shows a spectrum in which all S is
singly ionized.
The S$^+$ column density through the cell in the lower panel is similar
to the S$^+$ column density in the DLA shown in the upper panel.
\label{fig_abs}
}
\end{figure}

S II FUV lines are commonly observed in absorption in the interstellar and intergalactic media,
and can be used to probe the composition of the intervening clouds.
The lower panel of Figure \ref{fig_abs} shows 
the absorption spectrum of a pure S$^+$ gas with a column density of
$N($S$^+) = 10^{15.2} \pscm$.
We consider models of DLA clouds in the following section, but
present the absorption spectrum of a pure S$^+$ gas here for completeness. 
The $N($S$^+)$ column density was chosen to be representative of the column
density through the low-metallicity DLA clouds described in the next section. 
The Figure is limited to wavelengths $\lambda > 1000$\AA\ since shorter regions
are likely to be blocked by Lyman limit confusion.

Only a few S II lines (three lines near 1250 \AA) are present in the spectral region shown in Figure \ref{fig_abs}.
Other strong S II absorption lines are present, however.
Table \ref{tab_observations} lists all predicted S II lines with optical depths
greater than $10^{-3}$.
Most of the lines are in the EUV, making them unobservable at cosmological distances; 
but we note that there are two multiplets near 1021 \AA\ and 1204 \AA\ which
would be observable in higher column-density objects.

%%%%%%%%%%%%%%%%%%%%%%%%%%%%%%%%%%%%%%%%%%%%%%%%
\renewcommand{\baselinestretch}{0.75}
\begin{deluxetable}{lrl}
\tabletypesize{\footnotesize}
\tablecolumns{3} 
\tablewidth{0pt}
\tablecaption{
\label{tab_observations}
 S II mean optical depths for S$^+$-absorbing cloud in Fig.\ref{fig_abs}
} 
\tablehead{ 
\colhead{Ion} &
\colhead{$\lambda$(\AA)} & 
\colhead{Mean optical depth}
}
\startdata 
S II&  538& 1.18E$+$00\\
S II&  538& 9.32E$+$00\\
S II&  541& 1.39E$+$01\\
S II&  542& 1.71E$+$00\\
S II&  546& 2.22E$+$01\\
S II&  547& 2.61E$+$00\\
S II&  554& 3.86E$+$01\\
S II&  555& 4.29E$+$00\\
S II&  566& 7.60E$+$01\\
S II&  569& 7.81E$+$00\\
S II&  587& 1.81E$+$02\\
S II&  595& 1.71E$+$01\\
S II&  632& 6.12E$+$02\\
S II&  654& 6.77E$+$01\\
S II&  750& 3.27E$+$03\\
S II&  888& 6.25E$+$02\\
S II&  910& 2.35E$+$02\\
S II&  912& 1.17E$+$02\\
S II&	1021& 1.50E$-$02\\
S II&	1021& 1.56E$-$03\\
S II&	1204& 1.34E$-$03\\
S II& 1250& 1.21E$+$01\\
S II& 1253& 2.44E$+$01\\
S II& 1259& 3.72E$+$01\\
\enddata 
\end{deluxetable}

\subsection{Application to DLAs}

One aim of DLA absorption line spectroscopy is to be able to measure elemental abundances.
In the case of S, we are often limited to the S II lines described in this paper.
These can be used to infer the S$^+$ column density, but
to get abundances we must estimate the ionization fraction ratio S$^+$/S.

\citet{Howk99} pointed out that certain ion ratios can be used to estimate ionization
fractions of elements where only one stage of ionization is seen.
We redo a calculation in the spirit of theirs.
Like \citet{Howk99}, the SED is from \citet{HM96}
and the total neutral hydrogen column density of the cloud is taken as 
$N({\rm H}^0) = 10^{21} \pscm$,
roughly in the middle of the range of DLAs.
We assume a redshift of $z = 2$.
We assume ISM gas-phase abundances and dust with the metallicity
and dust to gas ratio reduced by 1 dex, as is typical of these objects.

Given these assumptions, the only free parameter is the gas density.
The metagalactic radiation background is assumed to be the only source of ionization.
Given this SED the ionization of a DLA will be determined by its density, since
the impinging flux of photons is constant.  In this case, lower density
gas will have a high ionization parameter \citep{AGN3}. The 
density range was chosen to cover the range of ionization parameters shown
in \citet{Howk99}.

\begin{figure}
\includegraphics[scale=1.05]{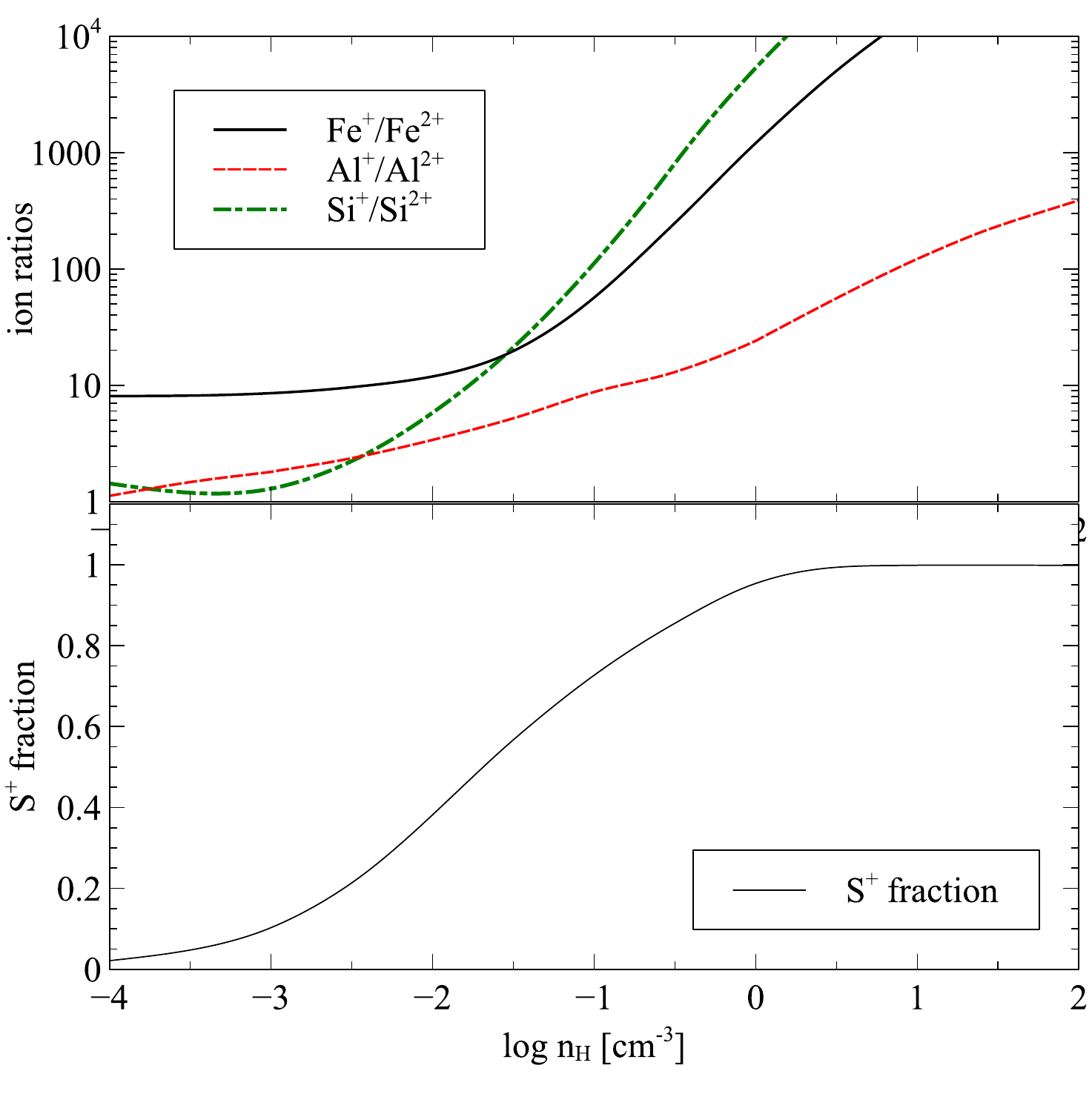}
\caption{ 
The upper panel shows the ratios of several commonly observed ions.
The lower panel shows the predicted S$^+$ ionization fraction.
Several ratios correlate with S$^+$ and can be used to estimate it.
\label{fig_IonFracs}
}
\end{figure}

Figure \ref{fig_IonFracs} shows the results.  The upper panel gives some
observable ion ratios while the lower panel shows the computed S$^+$ ionization fraction.
Clouds with densities greater than $\sim 1 \pcc$ will have nearly all S 
in the form of S$^+$.

The full DLA spectrum of a cloud with a hydrogen density of $1 \pcc$,
so that S$^+$ is the dominant ion stage, and
$N({\rm H}^0) = 10^{21} \pscm$, is shown in the upper panel of
Figure \ref{fig_abs} .
The S$^+$ column density in the lower panel was chosen to be similar
to the S$^+$ column density through the DLA shown in the upper panel.
These figures illustrate the potential of the new S II data for the purpose of 
making comparisons to observed spectra.

\section{Discussion}

Our oscillator strengths for most \ion{S}{2} absorption lines agree closely with 
previous calculations. For example, for the triplet at 1250.6, 1253.8, and 
1259.5 {\AA}, our calculations are lower than previous ones by 0.04 dex. Thus, 
the metallicities inferred from these lines would be higher by 0.04 dex. 
 It is reassuring that these metallicity corrections are relatively small. 

One surprising result of our work is that the uncertainties NIST placed on
the existing S II transition rates were too pessimistic.
Their quality flags indicate a typical uncertainty of roughly 30\%.
Our independent calculations confirm the predictions of previous work,
and suggest that the typical uncertainty is closer to 10\% for the strongest lines.

Our Cloudy simulations illustrate an astrophysical application of the atomic data 
calculations, and the resultant 
predictions for a large number of emission and 
absorption lines. Such predictions can be compared with observed line strengths 
to better constrain the properties of distant galaxies. Past S II absorption 
line observations  of DLAs have usually focussed on the strong lines at 
$\lambda \lambda 1250.58, 1253.81, 1259.52$. Our calculations confirm
that these are the strongest lines in the observable part of the spectrum. Our full S II dataset posted 
online contains many UV transitions, some of which could 
be detected in high-S/N spectra with future large telescopes. 

It is also important to estimate how much difference to the true [S/H] can be 
caused by ionization effects. Our Cloudy calculations in Fig. 10 indicate that 
the ionization correction for [S/H] derived from \ion{S}{2} / \ion{H}{1} is 
small for typical DLAs. We note, however, that these calculations are subject 
to uncertainties in recombination  coefficients. We plan to perform improved 
calculations of recombination coefficients in another part of this study. 

We also note that theoretically calculated transition wavelengths  can never reach the accuracy 
of experimental data. 
As is the standard practice in this field, we use experimental $\lambda$ values in our Cloudy 
simulation runs. 
Transition probabilities or oscillator strengths must be corrected for the difference
between experimental and theoretical energies, which is why our calculations
work in terms of the line strength $S$.
We do correct our transition rates by introducing
the experimental level energies to determine radiative transition parameters,
such as oscillator strengths $gf$ or transition probabilities $A$,
or tabulated data.

This  paper is a demonstration of the work we plan to do for other ions 
of S as well as for the observationally important ions of other elements from 
Al to Zn. The results of these broader calculations will be presented in 
several future papers and made available to the astrophysics community 
through incorporation into Cloudy. 

\acknowledgments

This work is supported by the National Science Foundation grant AST/1109061 
to Univ. of Kentucky and AST/1108830 to Univ. of South Carolina. 
VPK also acknowledges partial support from NSF (AST/0908890) and NASA 
(HST-GO-12536). GJF acknowledges additional support by NSF (AST/1108928), 
NASA (10-ATP10-0053, 10-ADAP10-0073, and NNX12AH73G), and STScI 
(HST-AR-12125.01, HST-AR- 13245, GO-12560, and HST-GO-12309).
RK's and PB's research is partially funded the European Social Fund under 
the Global Grant measure, project VP1-3.1-{\v S}MM-07-K-02-013.

\end{document}